\documentclass[twocolumn,aps,prl,showpacs]{revtex4-1}

\usepackage{graphicx}
\usepackage{amsmath,amssymb}
\usepackage{textcomp}
\usepackage{pdfpages}
\newcommand{\abs}[1]{\ensuremath{\left |{#1}\right |}}
\newcommand{\Rb}{\ensuremath{{}^{87}\mathrm{Rb}}}
\newcommand{\Erec}{\ensuremath{E_\mathrm{r}}}

\newcommand{\Lp}{\ensuremath{\mathcal{L}_+}}
\newcommand{\Lm}{\ensuremath{\mathcal{L}_-}}
\newcommand{\Lpm}{\ensuremath{\mathcal{L}_\pm}}
\newcommand{\C}{\ensuremath{\mathcal{C}}}
\newcommand{\xti}{\ensuremath{\xi_{i}}}
\newcommand{\avg}[1]{\ensuremath{\langle #1\rangle}}

\newcommand{\X}{\ensuremath{\hat{X}}}
\newcommand{\Gsc}{\Gamma_\mathrm{sc}}
\newcommand{\gc}{\gamma_\mathrm{c}}
\newcommand{\ket}[1]{\ensuremath{\vert #1 \rangle}}

\newcommand{\gmix}{\ensuremath{\gamma_\mathrm{m}}}
\newcommand{\gred}{\ensuremath{\gamma'_\mathrm{m}}}

\newcommand{\ncoll}{\ensuremath{\avg{n}}}
\newcommand{\nmin}{\ensuremath{\avg{n}_\mathrm{min}}}
\newcommand{\nbath}{\ensuremath{n_\mathrm{bath}}}

\newcommand{\Pcool}{\ensuremath{P_\mathrm{c}}}
\newcommand{\micro}[1]{\ensuremath{\mu\textrm{#1}}}
\newcommand{\gtot}{\ensuremath{\gamma_\mathrm{exp}}}

\newcommand{\Gom}{\ensuremath{\mathcal{G}}}
\newcommand{\Xzp}{\ensuremath{X_\mathrm{0}}}

\newcommand{\cavshift}{\ensuremath{\delta\omega_N}}

\begin{document}

\title{Optomechanical Cavity Cooling of an Atomic Ensemble}

\author{Monika H. Schleier-Smith}
\affiliation{
Department of Physics, MIT-Harvard Center for Ultracold Atoms,
and Research Laboratory of Electronics, Massachusetts Institute of Technology,
Cambridge, Massachusetts 02139, USA}

\author{Ian D. Leroux}
\affiliation{
Department of Physics, MIT-Harvard Center for Ultracold Atoms,
and Research Laboratory of Electronics, Massachusetts Institute of Technology,
Cambridge, Massachusetts 02139, USA}

\author{Hao Zhang}
\affiliation{
Department of Physics, MIT-Harvard Center for Ultracold Atoms,
and Research Laboratory of Electronics, Massachusetts Institute of Technology,
Cambridge, Massachusetts 02139, USA}

\author{Mackenzie A. Van Camp}
\affiliation{
Department of Physics, MIT-Harvard Center for Ultracold Atoms,
and Research Laboratory of Electronics, Massachusetts Institute of Technology,
Cambridge, Massachusetts 02139, USA}

\author{Vladan Vuleti\'{c}}
\affiliation{
Department of Physics, MIT-Harvard Center for Ultracold Atoms,
and Research Laboratory of Electronics, Massachusetts Institute of Technology,
Cambridge, Massachusetts 02139, USA}

\date{\today}
\begin{abstract}
We demonstrate cavity sideband cooling of a single collective motional mode of an atomic ensemble down to a mean phonon occupation number $\nmin = 2.0_{-0.3}^{+0.9}$.  Both $\nmin$ and the observed cooling rate are in good agreement with an optomechanical model.  The cooling rate constant is proportional to the total photon scattering rate by the ensemble, demonstrating the cooperative character of the light-emission-induced cooling process. We deduce fundamental limits to cavity-cooling either the collective mode or, sympathetically, the single-atom degrees of freedom.

\end{abstract}

\maketitle

Cavity cooling \cite{Horak97,Vuletic00,Vuletic01,Horak01,Domokos03} is unique among laser cooling techniques in that it is applicable, in principle, to arbitrary scatterers of light.  The energy spectrum of the scattered field---which governs the cooling dynamics and equilibrium temperature---is shaped by the cavity resonance rather than by the internal structure of the scatterer.  Cavity cooling thus offers enticing prospective applications, from preparing ultracold molecular gases \cite{Lev08,Morigi07} to continuous cooling of qubit registers with far-detuned light \cite{Griessner04}.  In experiments to date \cite{Maunz04,Leibrandt09,Chan03}, cavity cooling of one atom \cite{Maunz04} or ion \cite{Leibrandt09} is well described by a semiclassical model \cite{Vuletic00,Vuletic01}.  In the case of an ensemble, the coupling of many particles to a single cavity mode can yield nontrivial collective dynamics \cite{Nagorny03,*Kruse03,*Black03,Chan03,Murch08,Brennecke08,*Purdy10}, such as enhanced cooling of the center-of-mass motion \cite{Chan03}.

\begin{figure}
  \includegraphics[width=0.9\columnwidth]{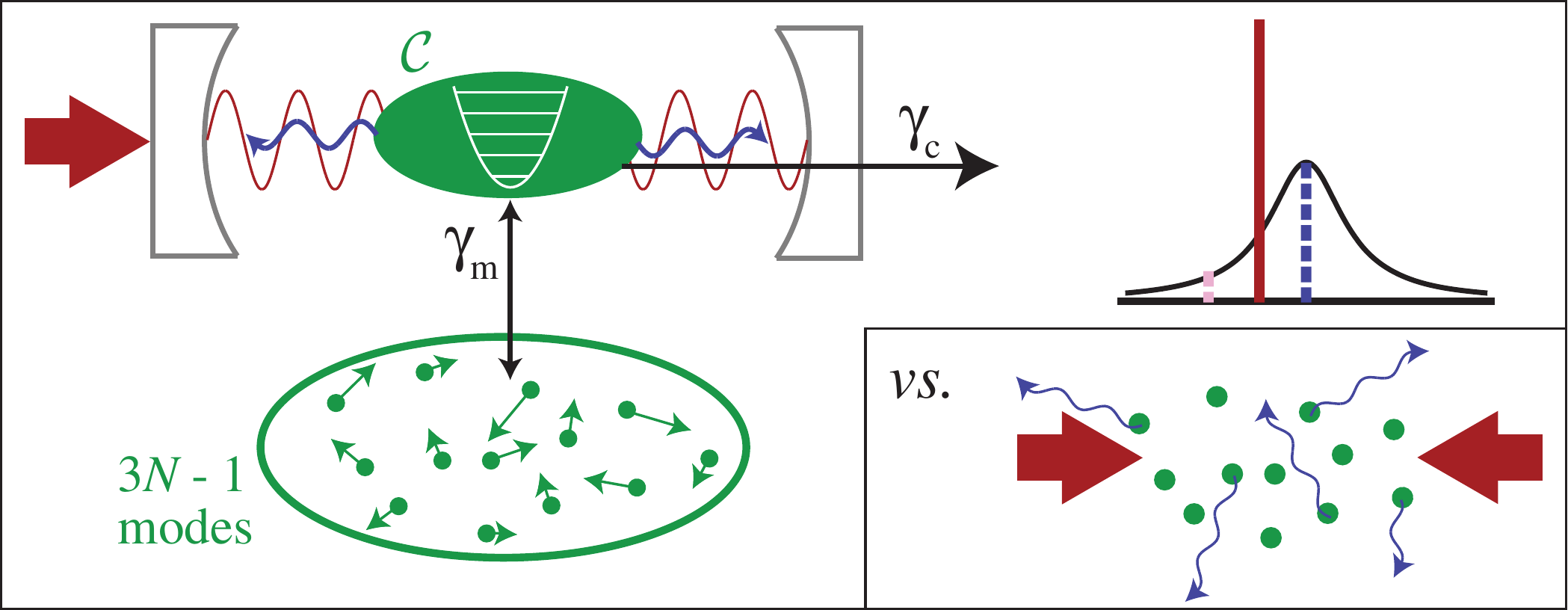}
  \caption{(Color online) Ensemble cavity cooling.  A probe laser (red) is placed at red detuning from cavity resonance to enhance anti-Stokes scattering into the cavity (blue), which cools a single collective mode $\C$ (solid green oval) at a cooperatively enhanced rate $\gamma_c$.  Single-particle modes can only be cooled by mixing (at rate $\gmix$) with $\C$.  This differs from ordinary laser cooling (inset), where free-space emission causes the atoms to be cooled independently.}\label{fig:setup}
\end{figure}

Ensemble cavity cooling (Fig. \ref{fig:setup}) differs markedly from conventional laser cooling, where emission into a plethora of free-space field modes allows for simultaneous and independent cooling of all atoms, or equivalently, all motional degrees of freedom of the ensemble.  In cavity cooling, a single collective motional mode $\C$ can be defined that is maximally coupled to the cavity \cite{Murch08}, while all other ensemble modes are decoupled from the cavity due to destructive interference in the light scattering from different atoms.  The coupling of $\C$ to the cavity is cooperatively enhanced by constructive interference in proportion to atom number \cite{Chan03,Lev08}, allowing $\C$ to be cooled faster---and to lower temperatures---than a single atom.

Pioneering experiments \cite{Brennecke08,Murch08,Purdy10} have recently demonstrated that the cavity-coupled collective mode $\C$ can be studied using the concepts of optomechanics \cite{Braginsky77}.  Indeed, the cooperative cooling of $\C$---in the limit of weak mixing with other ensemble modes---is equivalent to the single-mode cooling \cite{Braginsky77, Marquardt07, WilsonRae07, *Genes08} of macroscopic mechanical oscillators \cite{Kippenberg08, *Arcizet06, *Gigan06, *Schliesser06, *Thompson08,Rocheleau10} by radiation pressure.  Compared with solid-state mechanical oscillators, the collective atomic oscillator $\C$ inhabits a different parameter regime---of low mass and correspondingly large zero-point length---that may facilitate observing the quantization of mechanical energy \cite{Miao09}.  Furthermore, the internal degrees of freedom in an atomic ensemble constitute an extra tool for manipulating the motional quantum state.  The collective motion could, e.g., be squeezed by quantum state mapping from the ensemble spin \cite{Genes11}.

To cavity-cool the single-particle degrees of freedom in the ensemble, mixing between $\C$ and other motional modes may be introduced by an anharmonic or inhomogeneous trapping potential, or by collisions.  While such cooling has been the subject of significant theoretical studies, including detailed numerical modeling \cite{Horak01,Domokos03}, experiments confirming the predictions are few.

In this Letter, we cavity-cool and directly observe the relevant collective mode $\C$ of a trapped atomic ensemble.   The rate constant of the cooling depends linearly on both photon scattering rate per atom and atom number, demonstrating that the cooling relies on the cooperative emission of light by the ensemble.
Our results are well described by adapting an optomechanical model \cite{Marquardt07} to our system, where the mechanical oscillator $\C$ has a very small mass $M=(10^{-23}-10^{-21})$~kg, a frequency of 500 kHz (half the 1 MHz cavity linewidth), and a comparatively low quality factor $Q=19$.  We verify the agreement with optomechanical theory for a wide range of collective-mode occupation numbers up to $\ncoll \sim 10^3$, and we demonstrate cooling down to $\nmin=2.0_{-0.3}^{+0.9}$, close to the theoretical limit for our parameters.

The optomechanical interaction Hamiltonian $H$ in our system arises from a position-dependent dispersive coupling of the atoms to the cavity mode.  Formally, $H$ describes the dipole coupling of $N$ atoms with position operators $\hat{x}_i$ to light in a standing-wave cavity mode (``probe'' mode, with wavenumber $k$ and annihilation operator $\hat{a}$) at large detuning $\Delta$ from atomic resonance relative to the excited-state linewidth $\Gamma$.  Adiabatic elimination of the excited state yields $H = \hbar\Omega \sum_{i=1}^N \sin^2(k \hat{x}_i) \hat{a}^\dagger\hat{a}$, where $\Omega=g^2/\Delta$---with vacuum Rabi frequency $2g$---represents the dispersive shift of the cavity resonance due to a single atom at an antinode, or equivalently, the ac Stark shift experienced by such an atom per intracavity photon.  In our experiment, similar to Ref. \cite{Murch08}, the atoms are trapped along the cavity axis in an optical lattice incommensurate with the probe mode.  In the Lamb-Dicke regime, where the deviation $\tilde{x}_i\equiv\hat{x}_i-\xti$ of each atom from the local trap minimum at $\xti$ satisfies $\avg{(k\tilde{x}_i)^2}\ll 1$, the Hamiltonian $H$ can be written in terms of a single collective mode $\C$ of harmonic motion at the trap frequency $\omega_t$ \cite{Murch08}, with position operator $\X\equiv N^{-1} \sum_{i=1}^N \sin(2k \xti) \tilde{x}_i$ \cite{SM}. In terms of $\X$,
\begin{equation}\label{eq:H}
H = \hbar \Gom\X \hat{a}^\dagger\hat{a},
\end{equation}
where we have absorbed an overall shift $\cavshift \equiv \Omega \sum_i \sin^2(k \xti)$ into the cavity resonance frequency. Eq. \ref{eq:H} represents the canonical optomechanical interaction \cite{Braginsky77,Marquardt07,WilsonRae07,Genes08} describing an intensity-dependent force of strength $\hbar \Gom=N \hbar \Omega k$ per photon, or equivalently, a cavity frequency shift $\Gom\X$ proportional to $\X$.

For a probe laser detuned from the cavity line of width $\kappa$, small shifts $|\Gom\X|<\kappa$ yield proportional changes in intracavity and transmitted power.  The $\X$-dependent transmission can be used to monitor mode $\C$, while the $\X$-dependent changes in intracavity intensity---delayed by the cavity response---induce either cavity cooling or its reverse process, loosely termed cavity heating: specifically, the delay converts the position dependence into a velocity dependence of the force on the atoms, which either damps or coherently amplifies the collective motion depending on the sign of the laser-cavity detuning \cite{Vuletic00,Vuletic01}.

Viewed in the frequency domain, the dissipative process arises from unequal scattering rates on the Stokes and anti-Stokes sidebands due to the cavity resonance \cite{Vuletic01}.  The full optomechanical Hamiltonian \cite{Marquardt07}, with the interaction term given by Eq. \ref{eq:H}, predicts a cooling power
\begin{equation}\label{eq:Pcool}
\Pcool=N \Gsc\eta \Erec \zeta \left( \ncoll \abs{\Lp}^2- \left( \ncoll+1 \right) \abs{\Lm}^2 \right),
\end{equation}
for a mean occupation number $\ncoll$ of mode $\C$; here $\Gsc=\avg{a^\dagger a}\Gamma g^2/\Delta^2$ is the photon scattering rate of a single atom at a probe antinode into free space, $\eta=4g^2/(\kappa\Gamma)$ the cavity-to-free-space scattering ratio (single-atom cooperativity) \cite{Vuletic01}, $\Erec=\hbar^2 k^2/(2m)$ the recoil energy for atomic mass $m$, $\zeta =N^{-1} \sum_i \sin^2(2k \xti)$,  and $\Lpm^{-1} = 1\mp 2i(\delta\pm\omega_t)/\kappa$, where $\delta$ is the probe-cavity detuning. In our experiments, where the atomic cloud is long ($\approx 1$~mm) compared to the 5-$\micro{m}$ beat length between trap and probe light, $\zeta=1/2$. For $\omega_t\gtrsim\kappa/2$, the cooling rate is maximized by placing the anti-Stokes sideband on resonance, $\delta=-\omega_t$. Eq. \ref{eq:Pcool} indicates a collective rate constant $\gamma_c=d\Pcool/d(\ncoll \hbar \omega_t)$ that is proportional to $N$ due to cooperative scattering: the larger the ensemble, the faster $\C$ is cooled.

\begin{figure}
  \includegraphics[width=0.9\columnwidth]{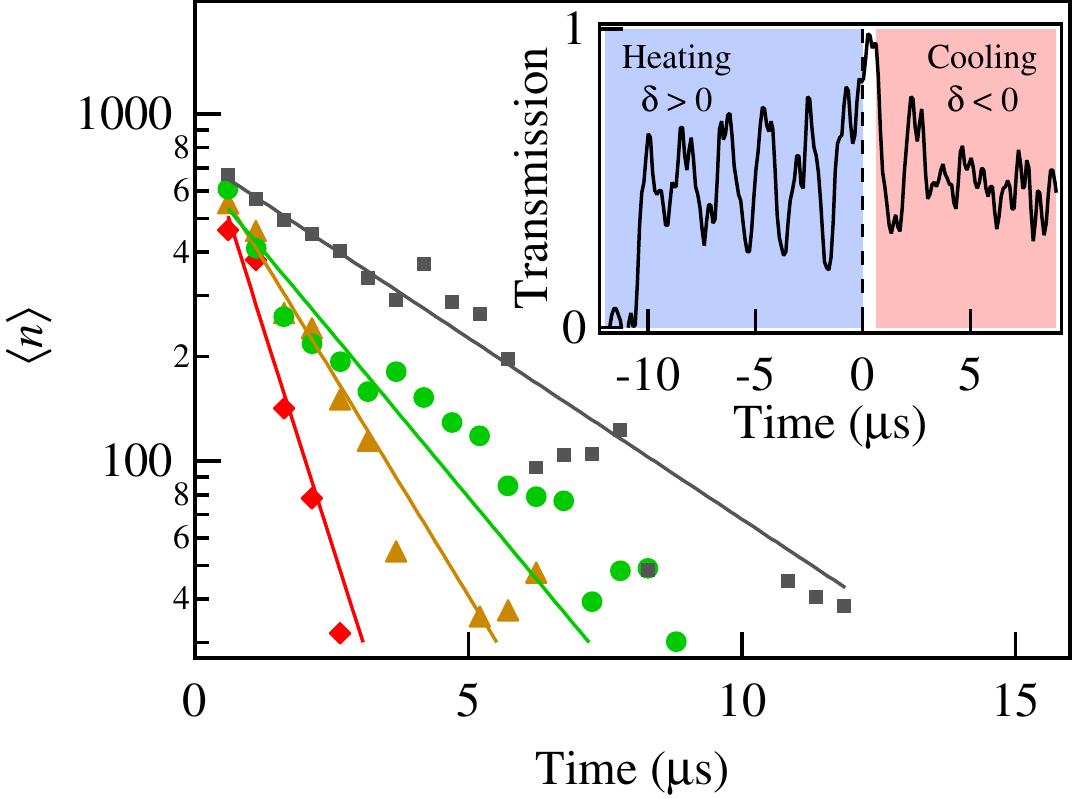}
  \caption{(Color online) Mean occupation number $\ncoll$ of mode $\C$ vs. time during cavity cooling at $\Gsc=1.1\times 10^5$~s$^{-1}$ (gray squares), $2.3\times 10^5$~s$^{-1}$ (green circles), $3.4\times 10^5$~s$^{-1}$ (gold triangles), and $6.4\times 10^5$~s$^{-1}$ (red diamonds).  Each dataset is obtained by averaging variances from 10 traces.  Inset: single trace of cavity transmission during cavity heating ($t<0$, blue background) followed by cooling ($t>0$, red background).}
\label{fig:ncoll}
\end{figure}

We study the cooling in a symmetric near-confocal optical cavity with linewidth $\kappa=2\pi\times 1.01(3)$~MHz at the wavelength $2\pi/k=$~780 nm of the $\Rb$ D$_2$ line, mode waist $w=56.9(4)~\micro{m}$, and cooperativity $\eta=0.203(7)$.  We trap $10^2$-$10^4$ atoms of $\Rb$ in the state $\ket{5 ^2S_{1/2},F=2,m_F=2}$ in the cavity mode in a standing wave of 851-nm light, with trap frequency $\omega_t/(2\pi) = 480(40)$~kHz and typical trap depth $U_0/h = 18(3)$~MHz.  A $\sigma^+$-polarized 780-nm probe laser drives the cavity on a TEM$_{00}$ mode at a detuning $\Delta/(2\pi) \ge 70$~MHz from the $\ket{5 ^2S_{1/2}, F=2}\rightarrow\ket{5 ^2P_{3/2}, F'=3}$ transition with linewidth $\Gamma=2\pi\times 6.1$~MHz. The atom number $N$ is measured via the average cavity shift $\cavshift$  \cite{SM}. To perform cavity cooling/heating, we detune the laser by $\delta=\mp \kappa/2\approx\mp\omega_t$ from cavity resonance, simultaneously probing the position $\X$ via the transmitted light. Note that we work with blue light-atom detuning $\Delta>0$, where free-space scattering results in Doppler heating.

We first verify cavity heating of mode $\C$ by choosing the probe-cavity detuning $\delta=+\kappa/2$.  Suddenly turning on the probe light triggers a collective oscillation that is rapidly amplified by parametric instability (inset to Fig. \ref{fig:ncoll}).  After typically 10~$\mu{s}$ of this heating, we switch to cavity cooling at $\delta=-\kappa/2$.  The mean occupation number $\ncoll$ of $\C$ is obtained from the observed time trace of the transmitted photon rate $R$ via the fractional variance $\sigma^2\equiv \overline{R^2}/\overline{R}^2-1$ in a sliding 2-$\mu{s}$ window.  The linear approximation $X \propto R-\overline{R}$ gives the relation $\sigma^2-\sigma^2_\mathrm{bg}=8(\Gom\Xzp/\kappa)^2 \abs{\Lp-\Lm}^2 \ncoll$, where $\Xzp=\sqrt{\hbar\zeta/(2Nm\omega_t)}$ and $\sigma^2_\mathrm{bg}$ is a constant technical-noise offset \cite{SM}. Fig. \ref{fig:ncoll} shows $\ncoll$ vs. time at four different probe powers, with fixed atom number $N=2800(400)$ and detuning $\Delta/(2\pi)=140$~MHz from atomic resonance.  The cooling is well described by an exponential decay with rate constant $\gtot$ that depends on the probe power. Consistent values of $\gtot$ are obtained by fitting an exponentially decaying sinusoid to the averaged transmission trace.

\begin{figure}
  \includegraphics[width=0.9\columnwidth]{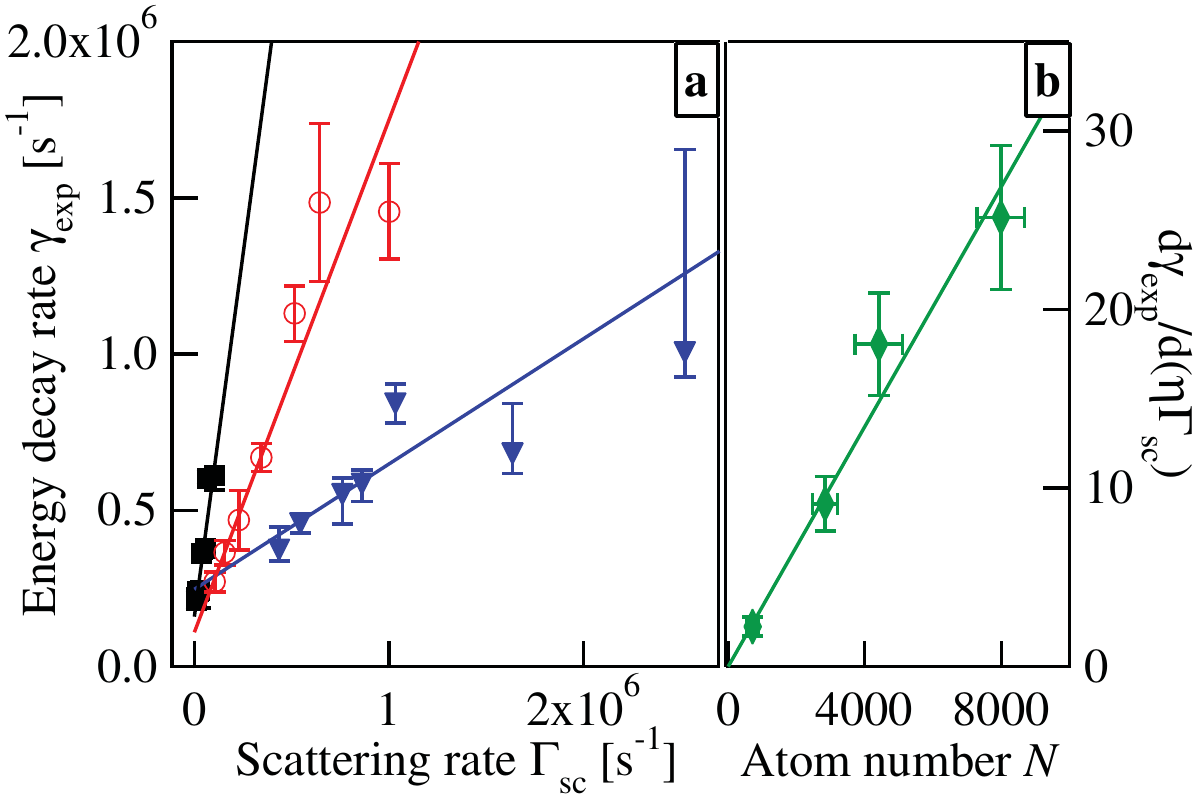}
  \caption{Collective cooling rates.  (a) Energy decay rate $\gtot$ vs. scattering rate $\Gsc$ for: $N$=8000(700), $\Delta/(2 \pi)=270$ MHz (solid black squares); $N$=2800(400), $\Delta/(2 \pi)=140$ MHz (open red circles); $N$=700(200), $\Delta/(2 \pi)=70$ MHz (solid blue triangles).  Lines are fits to data.  (b) $N$-dependence of cooling rate normalized to single-atom scattering rate into cavity.}\label{fig:cooling_rate}
\end{figure}

To compare $\gtot$ to the predicted cooling rate constant $\gamma_c$, we measure the dependence of $\gtot$ on the photon scattering rate $\Gsc=\overline{R}\eta \Gamma^2/(2\Delta^2)$ per atom into free space for various probe-atom detunings $\Delta$ and atom numbers $N$. As Fig. \ref{fig:cooling_rate}(a) shows, the data are consistent with a linear model $\gtot= f(N)\eta\Gsc +\gmix$. The offset $\gmix=1.6(6)\times 10^5$/s indicates a quality factor $Q=\omega_t/\gmix\approx 19$ for mode $\C$, largely attributable to mixing with other motional modes in the anharmonic trapping potential. Note that our system allows cavity cooling at very low $Q$ compared to solid mechanical oscillators \cite{Kippenberg08, Arcizet06, Gigan06, Schliesser06, Thompson08, Rocheleau10} because the ``thermal bath" comprising the other $3N-1$ ensemble modes has a sub-mK temperature.

To verify the cooperative nature of the cavity cooling of mode $\C$, we plot in Fig. \ref{fig:cooling_rate}(b) the fitted slopes $f(N)$ as displayed in \ref{fig:cooling_rate}(a) vs. atom number $N$. Accounting for the slight ($< 20 \%$) reduction of the cooperativity $\eta$ due to atomic absorption, the measured dependence $d\gtot/d(\eta\Gsc)=3.4(5)\times 10^{-3} N$ agrees well with the prediction from cavity cooling $\gc/(\eta\Gsc)= 3.0(2)\times 10^{-3} N$. This confirms that the collective-mode cooling speed increases linearly with ensemble size and is proportional to the total power scattered by the ensemble into the cavity.

\begin{figure}
  \includegraphics[width=0.9\columnwidth]{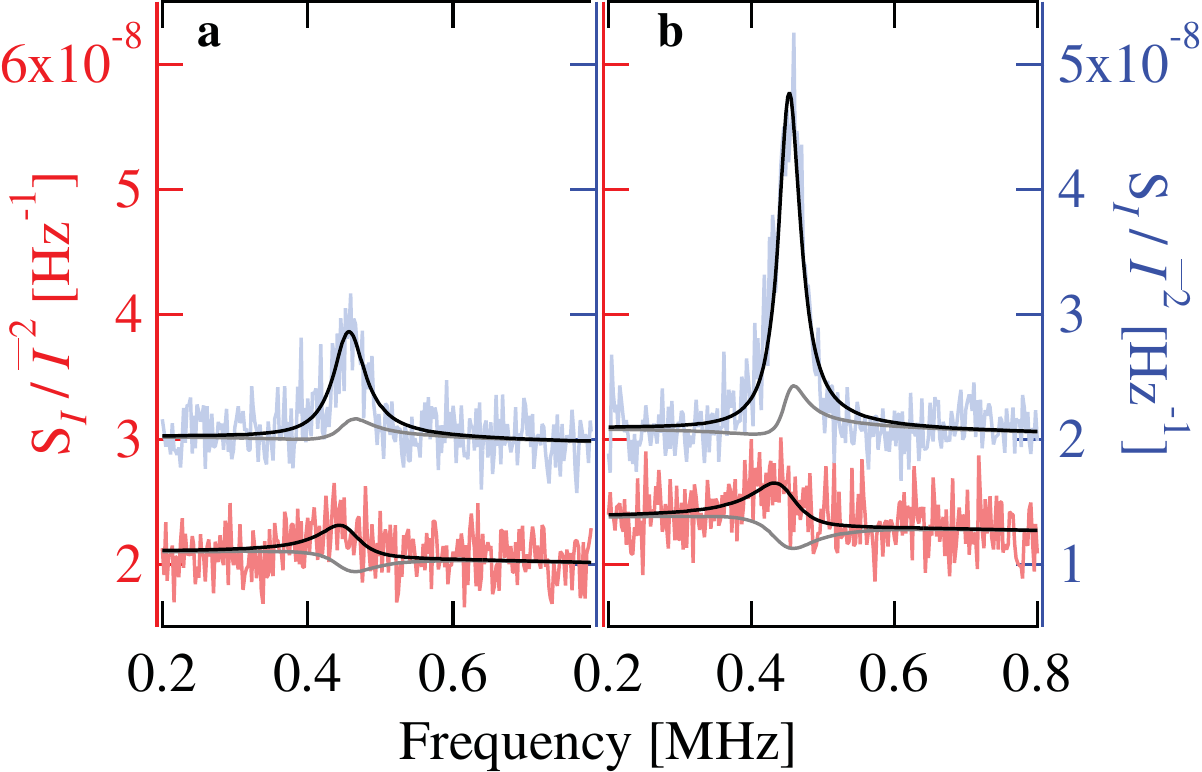}
  \caption{Spectra of fractional transmission fluctuations $S_{I}/\overline{I}^2$ taken with (a) N=230(50) atoms and (b) N=450(90) atoms during cavity cooling ($\delta=-\kappa/2$; red lines) or heating ($\delta=+\kappa/2$; blue lines).  Black curves are fits; subtraction of the background level $S_\mathrm{bg}$ (gray curves) yields collective mode occupation ${\ncoll}^\pm$ at $\delta=\pm\kappa/2$: (a) ${\ncoll}^+ = 4.4\pm 0.7$, ${\ncoll}^- = 2.3_{-0.3}^{+0.7}$; (b) ${\ncoll}^+ = 7\pm 1$, ${\ncoll}^- = 2.0_{-0.3}^{+0.9}$.}\label{fig:spectra}
\end{figure}

To determine the equilibrium temperature of $\C$ under cooling, we require---given our detection noise---a longer observation time than shown in Fig. \ref{fig:ncoll}.  We therefore observe the cooling or (for comparison) heating in spectra obtained from 150 time traces of the cavity transmission, each  $440$-$\mu{s}$ long, with $\overline{R} = 1.2(2)\times 10^9$~s$^{-1}$.  Fig. \ref{fig:spectra} shows normalized one-sided spectral densities $S_{I}/\overline{I}^2$ of photocurrent $I\propto R$ with (a) $N = 230(50)$ and (b) $N = 450(90)$ atoms at a detuning $\Delta/(2 \pi) = 70$ MHz from atomic resonance.  Each spectrum displays a peak at $\omega_t$ with an area approximately proportional to both atom number $N$ and mean occupation number $\ncoll$.  The disparity in area between cooling and heating increases with $N$ due to the cooperative nature of the processes.

We fit the spectra in Fig. \ref{fig:spectra} with a quantum mechanical model (black curves) adapted \cite{SM} from Ref. \cite{Marquardt07}. The model $S_I/\overline{I}^2 = S_\mathrm{mech} + S_\mathrm{bg}$ contains the signal $S_\mathrm{mech}\approx(2\Gom/\kappa)^2\abs{\Lp-\Lm}^2 S_{X}$ arising from atomic motion with spectral density $S_{X}$; and a background $S_\mathrm{bg}$ (gray curves) that is dominated by electronic photodetector noise but also accounts for photon shot noise, slightly smaller fluctuations from laser phase noise, and frequency-dependent correlations between light noise and atomic motion. These last are responsible for the dips in $S_\mathrm{bg}$ below the white noise \cite{Rocheleau10}.  With the photon rate $\overline{R}$ and optomechanical coupling $\Gom$ constrained to their independently measured and calculated values, the cooling spectra are well fit by taking the collective mode to be coupled to a white Markovian bath with $\avg{\nbath} = 3.1(4)$ motional quanta per mode; the corresponding coupling rate $\gred=2.6(1.1)\times 10^5$~s$^{-1}$ is consistent with the mixing rate $\gmix$ from Fig. \ref{fig:cooling_rate}.  Fits to the heating spectra, complicated by sympathetic heating of other modes, indicate a higher mixing rate of $4.8(5)\times 10^5$ /s.

The bath occupation is consistent with a measured upper bound on the axial temperature of $150(50)$~$\mu$K, corresponding to $\avg{\nbath} = 6(2)$ \cite{SM}.  The white spectrum of $\nbath$ is a simplistic ansatz but helps to establish the background $S_\mathrm{bg}$ and thus the motional spectrum $S_{X}$.  By subtracting $S_\mathrm{bg}$ from the measured spectrum, we obtain a minimum mean occupation number of $\C$ of $\nmin=2.0_{-0.3}^{+0.9}$ with $N=450(90)$ atoms.  Note that failing to account for the dip in $S_\mathrm{bg}$ would underestimate $\nmin$.

We now consider limits to cooling the collective mode.  For $\gamma_c\gg\gmix$, the cooling power $\Pcool\propto N\eta$ competes only with the $N$-independent recoil heating $P_\mathrm{rec}\approx\Erec\Gsc$ of $\C$, yielding a fundamental limit $\ncoll\ge n_0+D(1+n_0)/(N\eta)$, where $n_0\equiv(\kappa/4\omega_t)^2$ and $D$ is a prefactor of order unity \cite{Vuletic01}.  Thus, for large collective cooperativity $N\eta\gg 1$ (easy to achieve), the resolved-sideband regime $n_0<1$ in principle allows ground-state cooling \cite{Vuletic01,Marquardt07,WilsonRae07,Genes08} of $\C$. The thermal heat load from other modes mixing at rate $\gmix$ with $\C$ then sets the limit $\ncoll\gtrsim\avg{\nbath}\gmix/(\gmix+\gamma_c)$.  While this limit improves with increasing cooling rate $\gamma_c$, for the values $(\omega_t,\kappa,Q=\omega_t/\gred)$ in Fig. \ref{fig:spectra} amplification of low-frequency noise on approaching the regime of static bistability $\gamma_c\gtrsim\omega_t^2/\kappa$ \cite{Marquardt07} sets a bound $\ncoll \geq 1.5$, even though $n_0=0.3$.

A low occupation $\ncoll$ of $\C$ is disadvantageous for cooling the individual atoms, since the absolute cooling power is proportional to $\ncoll$ (see Eq. \ref{eq:Pcool}). Cooling of all degrees of freedom is thus facilitated by strong mixing $\gmix \gg \gamma_c$ that keeps $\C$ in thermal equilibrium with the other $3N-1$ modes.  The cooling power per atom $\Pcool/N$ then approaches that of an isolated atom.  Thus, even in an ensemble, recoil heating sets a limit for the temperature of individual atoms $\avg{n_i}\gtrsim n_0 + 1/\eta$ that depends on the \textit{single-atom} cooperativity $\eta$: ground-state cooling requires $\eta>1$ \cite{Vuletic01}.  Whether the same result holds in other cooling geometries, e.g., with transverse pumping \cite{Domokos03,Lev08,Griesser10}, is under investigation \cite{Griesser10}.

Even for $\eta<1$, ground-state cooling of $\C$ alone---in future experiments deeper in the resolved sideband regime---may enable the preparation of non-classical motional states \cite{Kippenberg08,Genes11}.  Further, the sensitive detection demonstrated here for $\X$ can alternatively be applied to measure $\X^2$ and thereby observe phonon shot noise \cite{Clerk10b} or perhaps even quantum jumps in $n$ \cite{Miao09}.

This work was supported in part by the NSF, DARPA, and the NSF Center for Ultracold Atoms.  M.~S.-S. acknowledges support from the Hertz Foundation and NSF, I.~D.~L. acknowledges support from NSERC, and M.~V.~C. acknowledges support from the NSF IGERT program.

\bibliography{cavcool}

\begin{thebibliography}{32}%
\makeatletter
\providecommand \@ifxundefined [1]{%
 \@ifx{#1\undefined}
}%
\providecommand \@ifnum [1]{%
 \ifnum #1\expandafter \@firstoftwo
 \else \expandafter \@secondoftwo
 \fi
}%
\providecommand \@ifx [1]{%
 \ifx #1\expandafter \@firstoftwo
 \else \expandafter \@secondoftwo
 \fi
}%
\providecommand \natexlab [1]{#1}%
\providecommand \enquote  [1]{``#1''}%
\providecommand \bibnamefont  [1]{#1}%
\providecommand \bibfnamefont [1]{#1}%
\providecommand \citenamefont [1]{#1}%
\providecommand \href@noop [0]{\@secondoftwo}%
\providecommand \href [0]{\begingroup \@sanitize@url \@href}%
\providecommand \@href[1]{\@@startlink{#1}\@@href}%
\providecommand \@@href[1]{\endgroup#1\@@endlink}%
\providecommand \@sanitize@url [0]{\catcode `\\12\catcode `\$12\catcode
  `\&12\catcode `\#12\catcode `\^12\catcode `\_12\catcode `\%12\relax}%
\providecommand \@@startlink[1]{}%
\providecommand \@@endlink[0]{}%
\providecommand \url  [0]{\begingroup\@sanitize@url \@url }%
\providecommand \@url [1]{\endgroup\@href {#1}{\urlprefix }}%
\providecommand \urlprefix  [0]{URL }%
\providecommand \Eprint [0]{\href }%
\providecommand \doibase [0]{http://dx.doi.org/}%
\providecommand \selectlanguage [0]{\@gobble}%
\providecommand \bibinfo  [0]{\@secondoftwo}%
\providecommand \bibfield  [0]{\@secondoftwo}%
\providecommand \translation [1]{[#1]}%
\providecommand \BibitemOpen [0]{}%
\providecommand \bibitemStop [0]{}%
\providecommand \bibitemNoStop [0]{.\EOS\space}%
\providecommand \EOS [0]{\spacefactor3000\relax}%
\providecommand \BibitemShut  [1]{\csname bibitem#1\endcsname}%
\let\auto@bib@innerbib\@empty
\bibitem [{\citenamefont {Horak}\ \emph {et~al.}(1997)\citenamefont {Horak},
  \citenamefont {Hechenblaikner}, \citenamefont {Gheri}, \citenamefont
  {Stecher},\ and\ \citenamefont {Ritsch}}]{Horak97}%
  \BibitemOpen
  \bibfield  {author} {\bibinfo {author} {\bibfnamefont {P.}~\bibnamefont
  {Horak}}, \bibinfo {author} {\bibfnamefont {G.}~\bibnamefont
  {Hechenblaikner}}, \bibinfo {author} {\bibfnamefont {K.~M.}\ \bibnamefont
  {Gheri}}, \bibinfo {author} {\bibfnamefont {H.}~\bibnamefont {Stecher}}, \
  and\ \bibinfo {author} {\bibfnamefont {H.}~\bibnamefont {Ritsch}},\
  }\href@noop {} {\bibfield  {journal} {\bibinfo  {journal} {Phys. Rev. Lett.}\
  }\textbf {\bibinfo {volume} {79}},\ \bibinfo {pages} {4974} (\bibinfo {year}
  {1997})}\BibitemShut {NoStop}%
\bibitem [{\citenamefont {Vuleti\ifmmode~\acute{c}\else \'{c}\fi{}}\ and\
  \citenamefont {Chu}(2000)}]{Vuletic00}%
  \BibitemOpen
  \bibfield  {author} {\bibinfo {author} {\bibfnamefont {V.}~\bibnamefont
  {Vuleti\ifmmode~\acute{c}\else \'{c}\fi{}}}\ and\ \bibinfo {author}
  {\bibfnamefont {S.}~\bibnamefont {Chu}},\ }\href@noop {} {\bibfield
  {journal} {\bibinfo  {journal} {Phys. Rev. Lett.}\ }\textbf {\bibinfo
  {volume} {84}},\ \bibinfo {pages} {3787} (\bibinfo {year}
  {2000})}\BibitemShut {NoStop}%
\bibitem [{\citenamefont {Vuleti\ifmmode~\acute{c}\else \'{c}\fi{}}\ \emph
  {et~al.}(2001)\citenamefont {Vuleti\ifmmode~\acute{c}\else \'{c}\fi{}},
  \citenamefont {Chan},\ and\ \citenamefont {Black}}]{Vuletic01}%
  \BibitemOpen
  \bibfield  {author} {\bibinfo {author} {\bibfnamefont {V.}~\bibnamefont
  {Vuleti\ifmmode~\acute{c}\else \'{c}\fi{}}}, \bibinfo {author} {\bibfnamefont
  {H.~W.}\ \bibnamefont {Chan}}, \ and\ \bibinfo {author} {\bibfnamefont
  {A.~T.}\ \bibnamefont {Black}},\ }\href@noop {} {\bibfield  {journal}
  {\bibinfo  {journal} {Phys. Rev. A}\ }\textbf {\bibinfo {volume} {64}},\
  \bibinfo {pages} {033405} (\bibinfo {year} {2001})}\BibitemShut {NoStop}%
\bibitem [{\citenamefont {Horak}\ and\ \citenamefont {Ritsch}(2001)}]{Horak01}%
  \BibitemOpen
  \bibfield  {author} {\bibinfo {author} {\bibfnamefont {P.}~\bibnamefont
  {Horak}}\ and\ \bibinfo {author} {\bibfnamefont {H.}~\bibnamefont {Ritsch}},\
  }\href@noop {} {\bibfield  {journal} {\bibinfo  {journal} {Phys. Rev. A}\
  }\textbf {\bibinfo {volume} {64}},\ \bibinfo {pages} {033422} (\bibinfo
  {year} {2001})}\BibitemShut {NoStop}%
\bibitem [{\citenamefont {Domokos}\ and\ \citenamefont
  {Ritsch}(2003)}]{Domokos03}%
  \BibitemOpen
  \bibfield  {author} {\bibinfo {author} {\bibfnamefont {P.}~\bibnamefont
  {Domokos}}\ and\ \bibinfo {author} {\bibfnamefont {H.}~\bibnamefont
  {Ritsch}},\ }\href@noop {} {\bibfield  {journal} {\bibinfo  {journal} {J.
  Opt. Soc. Am. B}\ }\textbf {\bibinfo {volume} {20}},\ \bibinfo {pages} {1098}
  (\bibinfo {year} {2003})}\BibitemShut {NoStop}%
\bibitem [{\citenamefont {Lev}\ \emph {et~al.}(2008)\citenamefont {Lev},
  \citenamefont {Vukics}, \citenamefont {Hudson}, \citenamefont {Sawyer},
  \citenamefont {Domokos}, \citenamefont {Ritsch},\ and\ \citenamefont
  {Ye}}]{Lev08}%
  \BibitemOpen
  \bibfield  {author} {\bibinfo {author} {\bibfnamefont {B.~L.}\ \bibnamefont
  {Lev}}, \bibinfo {author} {\bibfnamefont {A.}~\bibnamefont {Vukics}},
  \bibinfo {author} {\bibfnamefont {E.~R.}\ \bibnamefont {Hudson}}, \bibinfo
  {author} {\bibfnamefont {B.~C.}\ \bibnamefont {Sawyer}}, \bibinfo {author}
  {\bibfnamefont {P.}~\bibnamefont {Domokos}}, \bibinfo {author} {\bibfnamefont
  {H.}~\bibnamefont {Ritsch}}, \ and\ \bibinfo {author} {\bibfnamefont
  {J.}~\bibnamefont {Ye}},\ }\href@noop {} {\bibfield  {journal} {\bibinfo
  {journal} {Phys. Rev. A}\ }\textbf {\bibinfo {volume} {77}},\ \bibinfo
  {pages} {023402} (\bibinfo {year} {2008})}\BibitemShut {NoStop}%
\bibitem [{\citenamefont {Morigi}\ \emph {et~al.}(2007)\citenamefont {Morigi},
  \citenamefont {Pinkse}, \citenamefont {Kowalewski},\ and\ \citenamefont
  {de~Vivie-Riedle}}]{Morigi07}%
  \BibitemOpen
  \bibfield  {author} {\bibinfo {author} {\bibfnamefont {G.}~\bibnamefont
  {Morigi}}, \bibinfo {author} {\bibfnamefont {P.~W.~H.}\ \bibnamefont
  {Pinkse}}, \bibinfo {author} {\bibfnamefont {M.}~\bibnamefont {Kowalewski}},
  \ and\ \bibinfo {author} {\bibfnamefont {R.}~\bibnamefont
  {de~Vivie-Riedle}},\ }\href@noop {} {\bibfield  {journal} {\bibinfo
  {journal} {Phys. Rev. Lett.}\ }\textbf {\bibinfo {volume} {99}},\ \bibinfo
  {pages} {073001} (\bibinfo {year} {2007})}\BibitemShut {NoStop}%
\bibitem [{\citenamefont {Griessner}\ \emph {et~al.}(2004)\citenamefont
  {Griessner}, \citenamefont {Jaksch},\ and\ \citenamefont
  {Zoller}}]{Griessner04}%
  \BibitemOpen
  \bibfield  {author} {\bibinfo {author} {\bibfnamefont {A.}~\bibnamefont
  {Griessner}}, \bibinfo {author} {\bibfnamefont {D.}~\bibnamefont {Jaksch}}, \
  and\ \bibinfo {author} {\bibfnamefont {P.}~\bibnamefont {Zoller}},\
  }\href@noop {} {\bibfield  {journal} {\bibinfo  {journal} {Journal of Physics
  B: Atomic, Molecular and Optical Physics}\ }\textbf {\bibinfo {volume}
  {37}},\ \bibinfo {pages} {1419} (\bibinfo {year} {2004})}\BibitemShut
  {NoStop}%
\bibitem [{\citenamefont {Maunz}\ \emph {et~al.}(2004)\citenamefont {Maunz},
  \citenamefont {Puppe}, \citenamefont {Schuster}, \citenamefont {Syassen},
  \citenamefont {Pinkse},\ and\ \citenamefont {Rempe}}]{Maunz04}%
  \BibitemOpen
  \bibfield  {author} {\bibinfo {author} {\bibfnamefont {P.}~\bibnamefont
  {Maunz}}, \bibinfo {author} {\bibfnamefont {T.}~\bibnamefont {Puppe}},
  \bibinfo {author} {\bibfnamefont {I.}~\bibnamefont {Schuster}}, \bibinfo
  {author} {\bibfnamefont {N.}~\bibnamefont {Syassen}}, \bibinfo {author}
  {\bibfnamefont {P.~W.~H.}\ \bibnamefont {Pinkse}}, \ and\ \bibinfo {author}
  {\bibfnamefont {G.}~\bibnamefont {Rempe}},\ }\href@noop {} {\bibfield
  {journal} {\bibinfo  {journal} {Nature}\ }\textbf {\bibinfo {volume} {428}},\
  \bibinfo {pages} {50} (\bibinfo {year} {2004})}\BibitemShut {NoStop}%
\bibitem [{\citenamefont {Leibrandt}\ \emph {et~al.}(2009)\citenamefont
  {Leibrandt}, \citenamefont {Labaziewicz}, \citenamefont
  {Vuleti\ifmmode~\acute{c}\else \'{c}\fi{}},\ and\ \citenamefont
  {Chuang}}]{Leibrandt09}%
  \BibitemOpen
  \bibfield  {author} {\bibinfo {author} {\bibfnamefont {D.~R.}\ \bibnamefont
  {Leibrandt}}, \bibinfo {author} {\bibfnamefont {J.}~\bibnamefont
  {Labaziewicz}}, \bibinfo {author} {\bibfnamefont {V.}~\bibnamefont
  {Vuleti\ifmmode~\acute{c}\else \'{c}\fi{}}}, \ and\ \bibinfo {author}
  {\bibfnamefont {I.~L.}\ \bibnamefont {Chuang}},\ }\href@noop {} {\bibfield
  {journal} {\bibinfo  {journal} {Phys. Rev. Lett.}\ }\textbf {\bibinfo
  {volume} {103}},\ \bibinfo {pages} {103001} (\bibinfo {year}
  {2009})}\BibitemShut {NoStop}%
\bibitem [{\citenamefont {Chan}\ \emph {et~al.}(2003)\citenamefont {Chan},
  \citenamefont {Black},\ and\ \citenamefont {Vuleti\ifmmode~\acute{c}\else
  \'{c}\fi{}}}]{Chan03}%
  \BibitemOpen
  \bibfield  {author} {\bibinfo {author} {\bibfnamefont {H.~W.}\ \bibnamefont
  {Chan}}, \bibinfo {author} {\bibfnamefont {A.~T.}\ \bibnamefont {Black}}, \
  and\ \bibinfo {author} {\bibfnamefont {V.}~\bibnamefont
  {Vuleti\ifmmode~\acute{c}\else \'{c}\fi{}}},\ }\href@noop {} {\bibfield
  {journal} {\bibinfo  {journal} {Phys. Rev. Lett.}\ }\textbf {\bibinfo
  {volume} {90}},\ \bibinfo {pages} {063003} (\bibinfo {year}
  {2003})}\BibitemShut {NoStop}%
\bibitem [{\citenamefont {Nagorny}\ \emph {et~al.}(2003)\citenamefont
  {Nagorny}, \citenamefont {Els\"asser},\ and\ \citenamefont
  {Hemmerich}}]{Nagorny03}%
  \BibitemOpen
  \bibfield  {author} {\bibinfo {author} {\bibfnamefont {B.}~\bibnamefont
  {Nagorny}}, \bibinfo {author} {\bibfnamefont {T.}~\bibnamefont {Els\"asser}},
  \ and\ \bibinfo {author} {\bibfnamefont {A.}~\bibnamefont {Hemmerich}},\
  }\href@noop {} {\bibfield  {journal} {\bibinfo  {journal} {Phys. Rev. Lett.}\
  }\textbf {\bibinfo {volume} {91}},\ \bibinfo {pages} {153003} (\bibinfo
  {year} {2003})}\BibitemShut {NoStop}%
\bibitem [{\citenamefont {Kruse}\ \emph {et~al.}(2003)\citenamefont {Kruse},
  \citenamefont {von Cube}, \citenamefont {Zimmermann},\ and\ \citenamefont
  {Courteille}}]{Kruse03}%
  \BibitemOpen
  \bibfield  {author} {\bibinfo {author} {\bibfnamefont {D.}~\bibnamefont
  {Kruse}}, \bibinfo {author} {\bibfnamefont {C.}~\bibnamefont {von Cube}},
  \bibinfo {author} {\bibfnamefont {C.}~\bibnamefont {Zimmermann}}, \ and\
  \bibinfo {author} {\bibfnamefont {P.~W.}\ \bibnamefont {Courteille}},\
  }\href@noop {} {\bibfield  {journal} {\bibinfo  {journal} {Phys. Rev. Lett.}\
  }\textbf {\bibinfo {volume} {91}},\ \bibinfo {pages} {183601} (\bibinfo
  {year} {2003})}\BibitemShut {NoStop}%
\bibitem [{\citenamefont {Black}\ \emph {et~al.}(2003)\citenamefont {Black},
  \citenamefont {Chan},\ and\ \citenamefont {Vuleti\ifmmode~\acute{c}\else
  \'{c}\fi{}}}]{Black03}%
  \BibitemOpen
  \bibfield  {author} {\bibinfo {author} {\bibfnamefont {A.~T.}\ \bibnamefont
  {Black}}, \bibinfo {author} {\bibfnamefont {H.~W.}\ \bibnamefont {Chan}}, \
  and\ \bibinfo {author} {\bibfnamefont {V.}~\bibnamefont
  {Vuleti\ifmmode~\acute{c}\else \'{c}\fi{}}},\ }\href@noop {} {\bibfield
  {journal} {\bibinfo  {journal} {Phys. Rev. Lett.}\ }\textbf {\bibinfo
  {volume} {91}},\ \bibinfo {pages} {203001} (\bibinfo {year}
  {2003})}\BibitemShut {NoStop}%
\bibitem [{\citenamefont {Murch}\ \emph {et~al.}(2008)\citenamefont {Murch},
  \citenamefont {Moore}, \citenamefont {Gupta},\ and\ \citenamefont
  {Stamper-Kurn}}]{Murch08}%
  \BibitemOpen
  \bibfield  {author} {\bibinfo {author} {\bibfnamefont {K.~W.}\ \bibnamefont
  {Murch}}, \bibinfo {author} {\bibfnamefont {K.~L.}\ \bibnamefont {Moore}},
  \bibinfo {author} {\bibfnamefont {S.}~\bibnamefont {Gupta}}, \ and\ \bibinfo
  {author} {\bibfnamefont {D.~M.}\ \bibnamefont {Stamper-Kurn}},\ }\href@noop
  {} {\bibfield  {journal} {\bibinfo  {journal} {Nature Physics}\ }\textbf
  {\bibinfo {volume} {4}},\ \bibinfo {pages} {561} (\bibinfo {year}
  {2008})}\BibitemShut {NoStop}%
\bibitem [{\citenamefont {Brennecke}\ \emph {et~al.}(2008)\citenamefont
  {Brennecke}, \citenamefont {Ritter}, \citenamefont {Donner},\ and\
  \citenamefont {Esslinger}}]{Brennecke08}%
  \BibitemOpen
  \bibfield  {author} {\bibinfo {author} {\bibfnamefont {F.}~\bibnamefont
  {Brennecke}}, \bibinfo {author} {\bibfnamefont {S.}~\bibnamefont {Ritter}},
  \bibinfo {author} {\bibfnamefont {T.}~\bibnamefont {Donner}}, \ and\ \bibinfo
  {author} {\bibfnamefont {T.}~\bibnamefont {Esslinger}},\ }\href@noop {}
  {\bibfield  {journal} {\bibinfo  {journal} {Science}\ }\textbf {\bibinfo
  {volume} {322}},\ \bibinfo {pages} {235} (\bibinfo {year}
  {2008})}\BibitemShut {NoStop}%
\bibitem [{\citenamefont {Purdy}\ \emph {et~al.}(2010)\citenamefont {Purdy},
  \citenamefont {Brooks}, \citenamefont {Botter}, \citenamefont {Brahms},
  \citenamefont {Ma},\ and\ \citenamefont {Stamper-Kurn}}]{Purdy10}%
  \BibitemOpen
  \bibfield  {author} {\bibinfo {author} {\bibfnamefont {T.~P.}\ \bibnamefont
  {Purdy}}, \bibinfo {author} {\bibfnamefont {D.~W.~C.}\ \bibnamefont
  {Brooks}}, \bibinfo {author} {\bibfnamefont {T.}~\bibnamefont {Botter}},
  \bibinfo {author} {\bibfnamefont {N.}~\bibnamefont {Brahms}}, \bibinfo
  {author} {\bibfnamefont {Z.-Y.}\ \bibnamefont {Ma}}, \ and\ \bibinfo {author}
  {\bibfnamefont {D.~M.}\ \bibnamefont {Stamper-Kurn}},\ }\href@noop {}
  {\bibfield  {journal} {\bibinfo  {journal} {Phys. Rev. Lett.}\ }\textbf
  {\bibinfo {volume} {105}},\ \bibinfo {pages} {133602} (\bibinfo {year}
  {2010})}\BibitemShut {NoStop}%
\bibitem [{\citenamefont {Braginsky}\ and\ \citenamefont
  {Manukin}(1977)}]{Braginsky77}%
  \BibitemOpen
  \bibfield  {author} {\bibinfo {author} {\bibfnamefont {V.~B.}\ \bibnamefont
  {Braginsky}}\ and\ \bibinfo {author} {\bibfnamefont {A.~B.}\ \bibnamefont
  {Manukin}},\ }\href@noop {} {\emph {\bibinfo {title} {Measurement of weak
  forces in physics experiments}}}\ (\bibinfo  {publisher} {University of
  Chicago Press},\ \bibinfo {address} {Chicago},\ \bibinfo {year}
  {1977})\BibitemShut {NoStop}%
\bibitem [{\citenamefont {Marquardt}\ \emph {et~al.}(2007)\citenamefont
  {Marquardt}, \citenamefont {Chen}, \citenamefont {Clerk},\ and\ \citenamefont
  {Girvin}}]{Marquardt07}%
  \BibitemOpen
  \bibfield  {author} {\bibinfo {author} {\bibfnamefont {F.}~\bibnamefont
  {Marquardt}}, \bibinfo {author} {\bibfnamefont {J.~P.}\ \bibnamefont {Chen}},
  \bibinfo {author} {\bibfnamefont {A.~A.}\ \bibnamefont {Clerk}}, \ and\
  \bibinfo {author} {\bibfnamefont {S.~M.}\ \bibnamefont {Girvin}},\
  }\href@noop {} {\bibfield  {journal} {\bibinfo  {journal} {Phys. Rev. Lett.}\
  }\textbf {\bibinfo {volume} {99}},\ \bibinfo {pages} {093902} (\bibinfo
  {year} {2007})}\BibitemShut {NoStop}%
\bibitem [{\citenamefont {Wilson-Rae}\ \emph {et~al.}(2007)\citenamefont
  {Wilson-Rae}, \citenamefont {Nooshi}, \citenamefont {Zwerger},\ and\
  \citenamefont {Kippenberg}}]{WilsonRae07}%
  \BibitemOpen
  \bibfield  {author} {\bibinfo {author} {\bibfnamefont {I.}~\bibnamefont
  {Wilson-Rae}}, \bibinfo {author} {\bibfnamefont {N.}~\bibnamefont {Nooshi}},
  \bibinfo {author} {\bibfnamefont {W.}~\bibnamefont {Zwerger}}, \ and\
  \bibinfo {author} {\bibfnamefont {T.~J.}\ \bibnamefont {Kippenberg}},\
  }\href@noop {} {\bibfield  {journal} {\bibinfo  {journal} {Phys. Rev. Lett.}\
  }\textbf {\bibinfo {volume} {99}},\ \bibinfo {pages} {093901} (\bibinfo
  {year} {2007})}\BibitemShut {NoStop}%
\bibitem [{\citenamefont {Genes}\ \emph {et~al.}(2008)\citenamefont {Genes},
  \citenamefont {Vitali}, \citenamefont {Tombesi}, \citenamefont {Gigan},\ and\
  \citenamefont {Aspelmeyer}}]{Genes08}%
  \BibitemOpen
  \bibfield  {author} {\bibinfo {author} {\bibfnamefont {C.}~\bibnamefont
  {Genes}}, \bibinfo {author} {\bibfnamefont {D.}~\bibnamefont {Vitali}},
  \bibinfo {author} {\bibfnamefont {P.}~\bibnamefont {Tombesi}}, \bibinfo
  {author} {\bibfnamefont {S.}~\bibnamefont {Gigan}}, \ and\ \bibinfo {author}
  {\bibfnamefont {M.}~\bibnamefont {Aspelmeyer}},\ }\href@noop {} {\bibfield
  {journal} {\bibinfo  {journal} {Phys. Rev. A}\ }\textbf {\bibinfo {volume}
  {77}},\ \bibinfo {pages} {033804} (\bibinfo {year} {2008})}\BibitemShut
  {NoStop}%
\bibitem [{\citenamefont {Kippenberg}\ and\ \citenamefont
  {Vahala}(2008)}]{Kippenberg08}%
  \BibitemOpen
  \bibfield  {author} {\bibinfo {author} {\bibfnamefont {T.~J.}\ \bibnamefont
  {Kippenberg}}\ and\ \bibinfo {author} {\bibfnamefont {K.~J.}\ \bibnamefont
  {Vahala}},\ }\href@noop {} {\bibfield  {journal} {\bibinfo  {journal}
  {Science}\ }\textbf {\bibinfo {volume} {321}},\ \bibinfo {pages} {1172}
  (\bibinfo {year} {2008})}\BibitemShut {NoStop}%
\bibitem [{\citenamefont {Arcizet}\ \emph {et~al.}(2006)\citenamefont
  {Arcizet}, \citenamefont {Cohadon}, \citenamefont {Briant}, \citenamefont
  {Pinard},\ and\ \citenamefont {Heidmann}}]{Arcizet06}%
  \BibitemOpen
  \bibfield  {author} {\bibinfo {author} {\bibfnamefont {O.}~\bibnamefont
  {Arcizet}}, \bibinfo {author} {\bibfnamefont {P.~F.}\ \bibnamefont
  {Cohadon}}, \bibinfo {author} {\bibfnamefont {T.}~\bibnamefont {Briant}},
  \bibinfo {author} {\bibfnamefont {M.}~\bibnamefont {Pinard}}, \ and\ \bibinfo
  {author} {\bibfnamefont {A.}~\bibnamefont {Heidmann}},\ }\href@noop {}
  {\bibfield  {journal} {\bibinfo  {journal} {Nature}\ }\textbf {\bibinfo
  {volume} {444}},\ \bibinfo {pages} {71} (\bibinfo {year} {2006})}\BibitemShut
  {NoStop}%
\bibitem [{\citenamefont {Gigan}\ \emph {et~al.}(2006)\citenamefont {Gigan},
  \citenamefont {Bohm}, \citenamefont {Paternostro}, \citenamefont {Blaser},
  \citenamefont {Langer}, \citenamefont {Hertzberg}, \citenamefont {Schwab},
  \citenamefont {Bauerle}, \citenamefont {Aspelmeyer},\ and\ \citenamefont
  {Zeilinger}}]{Gigan06}%
  \BibitemOpen
  \bibfield  {author} {\bibinfo {author} {\bibfnamefont {S.}~\bibnamefont
  {Gigan}}, \bibinfo {author} {\bibfnamefont {H.~R.}\ \bibnamefont {Bohm}},
  \bibinfo {author} {\bibfnamefont {M.}~\bibnamefont {Paternostro}}, \bibinfo
  {author} {\bibfnamefont {F.}~\bibnamefont {Blaser}}, \bibinfo {author}
  {\bibfnamefont {G.}~\bibnamefont {Langer}}, \bibinfo {author} {\bibfnamefont
  {J.~B.}\ \bibnamefont {Hertzberg}}, \bibinfo {author} {\bibfnamefont {K.~C.}\
  \bibnamefont {Schwab}}, \bibinfo {author} {\bibfnamefont {D.}~\bibnamefont
  {Bauerle}}, \bibinfo {author} {\bibfnamefont {M.}~\bibnamefont {Aspelmeyer}},
  \ and\ \bibinfo {author} {\bibfnamefont {A.}~\bibnamefont {Zeilinger}},\
  }\href@noop {} {\bibfield  {journal} {\bibinfo  {journal} {Nature}\ }\textbf
  {\bibinfo {volume} {444}},\ \bibinfo {pages} {67} (\bibinfo {year}
  {2006})}\BibitemShut {NoStop}%
\bibitem [{\citenamefont {Schliesser}\ \emph {et~al.}(2006)\citenamefont
  {Schliesser}, \citenamefont {Del'Haye}, \citenamefont {Nooshi}, \citenamefont
  {Vahala},\ and\ \citenamefont {Kippenberg}}]{Schliesser06}%
  \BibitemOpen
  \bibfield  {author} {\bibinfo {author} {\bibfnamefont {A.}~\bibnamefont
  {Schliesser}}, \bibinfo {author} {\bibfnamefont {P.}~\bibnamefont
  {Del'Haye}}, \bibinfo {author} {\bibfnamefont {N.}~\bibnamefont {Nooshi}},
  \bibinfo {author} {\bibfnamefont {K.~J.}\ \bibnamefont {Vahala}}, \ and\
  \bibinfo {author} {\bibfnamefont {T.~J.}\ \bibnamefont {Kippenberg}},\
  }\href@noop {} {\bibfield  {journal} {\bibinfo  {journal} {Phys. Rev. Lett.}\
  }\textbf {\bibinfo {volume} {97}},\ \bibinfo {pages} {243905} (\bibinfo
  {year} {2006})}\BibitemShut {NoStop}%
\bibitem [{\citenamefont {Thompson}\ \emph {et~al.}(2008)\citenamefont
  {Thompson}, \citenamefont {Zwickl}, \citenamefont {Jayich}, \citenamefont
  {Marquardt}, \citenamefont {Girvin},\ and\ \citenamefont
  {Harris}}]{Thompson08}%
  \BibitemOpen
  \bibfield  {author} {\bibinfo {author} {\bibfnamefont {J.~D.}\ \bibnamefont
  {Thompson}}, \bibinfo {author} {\bibfnamefont {B.~M.}\ \bibnamefont
  {Zwickl}}, \bibinfo {author} {\bibfnamefont {A.~M.}\ \bibnamefont {Jayich}},
  \bibinfo {author} {\bibfnamefont {F.}~\bibnamefont {Marquardt}}, \bibinfo
  {author} {\bibfnamefont {S.~M.}\ \bibnamefont {Girvin}}, \ and\ \bibinfo
  {author} {\bibfnamefont {J.~G.~E.}\ \bibnamefont {Harris}},\ }\href@noop {}
  {\bibfield  {journal} {\bibinfo  {journal} {Nature}\ }\textbf {\bibinfo
  {volume} {452}},\ \bibinfo {pages} {72} (\bibinfo {year} {2008})}\BibitemShut
  {NoStop}%
\bibitem [{\citenamefont {Rocheleau}\ \emph {et~al.}(2010)\citenamefont
  {Rocheleau}, \citenamefont {Ndukum}, \citenamefont {Macklin}, \citenamefont
  {Hertzberg}, \citenamefont {Clerk},\ and\ \citenamefont
  {Schwab}}]{Rocheleau10}%
  \BibitemOpen
  \bibfield  {author} {\bibinfo {author} {\bibfnamefont {T.}~\bibnamefont
  {Rocheleau}}, \bibinfo {author} {\bibfnamefont {T.}~\bibnamefont {Ndukum}},
  \bibinfo {author} {\bibfnamefont {C.}~\bibnamefont {Macklin}}, \bibinfo
  {author} {\bibfnamefont {J.~B.}\ \bibnamefont {Hertzberg}}, \bibinfo {author}
  {\bibfnamefont {A.~A.}\ \bibnamefont {Clerk}}, \ and\ \bibinfo {author}
  {\bibfnamefont {K.~C.}\ \bibnamefont {Schwab}},\ }\href@noop {} {\bibfield
  {journal} {\bibinfo  {journal} {Nature}\ }\textbf {\bibinfo {volume} {463}},\
  \bibinfo {pages} {72} (\bibinfo {year} {2010})}\BibitemShut {NoStop}%
\bibitem [{\citenamefont {Miao}\ \emph {et~al.}(2009)\citenamefont {Miao},
  \citenamefont {Danilishin}, \citenamefont {Corbitt},\ and\ \citenamefont
  {Chen}}]{Miao09}%
  \BibitemOpen
  \bibfield  {author} {\bibinfo {author} {\bibfnamefont {H.}~\bibnamefont
  {Miao}}, \bibinfo {author} {\bibfnamefont {S.}~\bibnamefont {Danilishin}},
  \bibinfo {author} {\bibfnamefont {T.}~\bibnamefont {Corbitt}}, \ and\
  \bibinfo {author} {\bibfnamefont {Y.}~\bibnamefont {Chen}},\ }\href@noop {}
  {\bibfield  {journal} {\bibinfo  {journal} {Phys. Rev. Lett.}\ }\textbf
  {\bibinfo {volume} {103}},\ \bibinfo {pages} {100402} (\bibinfo {year}
  {2009})}\BibitemShut {NoStop}%
\bibitem [{\citenamefont {Genes}\ \emph {et~al.}(2011)\citenamefont {Genes},
  \citenamefont {Ritsch}, \citenamefont {Drewsen},\ and\ \citenamefont
  {Dantan}}]{Genes11}%
  \BibitemOpen
  \bibfield  {author} {\bibinfo {author} {\bibfnamefont {C.}~\bibnamefont
  {Genes}}, \bibinfo {author} {\bibfnamefont {H.}~\bibnamefont {Ritsch}},
  \bibinfo {author} {\bibfnamefont {M.}~\bibnamefont {Drewsen}}, \ and\
  \bibinfo {author} {\bibfnamefont {A.}~\bibnamefont {Dantan}},\ }\href@noop {}
  {\  (\bibinfo {year} {2011})},\ \bibinfo {note} {arXiv:1105.0281v2
  [quant-ph]}\BibitemShut {NoStop}%
\bibitem [{SM()}]{SM}%
  \BibitemOpen
  \href@noop {} {}\bibinfo {note} {See EPAPS Document No. [number will be
  inserted by publisher] for further details and supporting experiments. For
  more information on EPAPS, see
  \url{http://www.aip.org/pubservs/epaps.html}.}\BibitemShut {Stop}%
\bibitem [{\citenamefont {{Griesser, T.}}\ \emph {et~al.}(2010)\citenamefont
  {{Griesser, T.}}, \citenamefont {{Ritsch, H.}}, \citenamefont {{Hemmerling,
  M.}},\ and\ \citenamefont {{Robb, G. R.M.}}}]{Griesser10}%
  \BibitemOpen
  \bibfield  {author} {\bibinfo {author} {\bibnamefont {{Griesser, T.}}},
  \bibinfo {author} {\bibnamefont {{Ritsch, H.}}}, \bibinfo {author}
  {\bibnamefont {{Hemmerling, M.}}}, \ and\ \bibinfo {author} {\bibnamefont
  {{Robb, G. R.M.}}},\ }\href@noop {} {\bibfield  {journal} {\bibinfo
  {journal} {Eur. Phys. J. D}\ }\textbf {\bibinfo {volume} {58}},\ \bibinfo
  {pages} {349} (\bibinfo {year} {2010})}\BibitemShut {NoStop}%
\bibitem [{\citenamefont {Clerk}\ \emph {et~al.}(2010)\citenamefont {Clerk},
  \citenamefont {Marquardt},\ and\ \citenamefont {Harris}}]{Clerk10b}%
  \BibitemOpen
  \bibfield  {author} {\bibinfo {author} {\bibfnamefont {A.~A.}\ \bibnamefont
  {Clerk}}, \bibinfo {author} {\bibfnamefont {F.}~\bibnamefont {Marquardt}}, \
  and\ \bibinfo {author} {\bibfnamefont {J.~G.~E.}\ \bibnamefont {Harris}},\
  }\href@noop {} {\bibfield  {journal} {\bibinfo  {journal} {Phys. Rev. Lett.}\
  }\textbf {\bibinfo {volume} {104}},\ \bibinfo {pages} {213603} (\bibinfo
  {year} {2010})}\BibitemShut {NoStop}%
\end{thebibliography}%
\clearpage


\section*{Supplementary material}

# Optomechanical Cavity Cooling of an Atomic Ensemble: Supplemental Material

Monika H. Schleier-Smith, Ian D. Leroux, Hao Zhang, Mackenzie A. Van Camp, and Vladan Vuletić
*Department of Physics, MIT-Harvard Center for Ultracold Atoms, and Research Laboratory of Electronics,*
*Massachusetts Institute of Technology, Cambridge, Massachusetts 02139, USA*

In this supplement, we begin in Sec. I by reviewing the description—previously derived by Murch *et al.* in Ref. [1]—of a trapped atomic ensemble as a single-mode mechanical oscillator coupled to the cavity. In Secs. II A-II D, we calculate the spectrum of intensity fluctuations at the output of the symmetric optical cavity, and we relate this spectrum to the spectrum of position fluctuations of the mechanical oscillator itself. We closely follow the approach of Marquardt *et al.* [2] for calculating the displacement spectrum of the mechanical oscillator, and we additionally include lowest-order effects of laser phase noise [3]. After accounting in Sec. II E for technical effects in our photodetection of the cavity transmission, we derive in Sec. II F the relation of measured photocurrent fluctuations to the occupation $\langle n \rangle$ of the collective mode. Our calibration of atom number is described in Sec. III. Finally, in Sec. IV we describe a measurement of the thermodynamic temperature associated with all $N$ modes of the ensemble's axial motion.

## I. HAMILTONIAN AND COLLECTIVE MODE

We first consider the Hamiltonian $H_{\text{sys}}$ of an ensemble of $N$ atoms that are harmonically trapped, with identical trap frequencies $\omega_t$, at various positions $\xi_i$ along the cavity axis. For dispersive coupling of the atoms to a cavity mode ("probe" mode) of frequency $\omega_0$ with annihilation operator $\hat{a}$, we have

$$H_{\text{sys}} = \sum_{i=1}^{N} \left[ \frac{1}{2} m \omega_t^2 (\hat{x}_i - \xi_i)^2 + \frac{\hat{p}_i^2}{2m} + \hbar \Omega \sin^2(k\hat{x}_i) \hat{a}^\dagger \hat{a} \right] + \hbar \omega_0 \hat{a}^\dagger \hat{a}. \tag{1}$$

Here, the atom-probe interaction is quantified by $\Omega = g^2/\Delta$, in terms of the vacuum Rabi frequency $2g$ and detuning $\Delta \gg \Gamma$ of the probe mode from the atomic resonance with linewidth $\Gamma$. In the Lamb-Dicke regime, where the motion $\tilde{x}_i \equiv \hat{x}_i - \xi_i \ll k^{-1}$ of each atom in its trap is small compared to the probe wavelength, the optomechanical interaction term can be expressed in terms of a single position variable $\hat{X} \equiv N^{-1} \sum_{i=1}^N \sin(2k\xi_i) \tilde{x}_i$ corresponding to the collective mode $\mathcal{C}$ discussed in the main text. The momentum conjugate to $\hat{X}$ is expressed in terms of the single-atom momenta $\hat{p}_i$ as $\hat{P} = N \sum_{i=1}^N \sin(2k\xi_i) \hat{p}_i / \sum_{i=1}^N \sin^2(2k\xi_i)$, such that $\hat{X}$ and $\hat{P}$ obey the canonical commutation relation $[\hat{X}, \hat{P}] = i\hbar$. To describe the full motion of the $N$-atom ensemble, one can construct an orthogonal basis comprising $\hat{X}$, $\hat{P}$, and additional coordinate pairs representing $3N-1$ other modes of ensemble motion with energy $H_\perp$; in this basis, the system Hamiltonian becomes

$$H_{\text{sys}} = \frac{1}{2} M \omega_t^2 \hat{X}^2 + \frac{\hat{P}^2}{2M} + \hbar \left( \omega_0 + \delta \omega_N + \mathcal{G} \hat{X} \right) \hat{a}^\dagger \hat{a} + H_\perp, \tag{2}$$

where $M = mN^2 / \sum_{i=1}^N \sin^2(2k\xi_i)$ represents the effective mass of the collective mode, $\delta\omega_N = \Omega \sum_{i=1}^N \sin^2(k\xi_i)$ is an overall shift of the cavity resonance due to the atoms, and $\mathcal{G} = N\Omega k$ is the change in this cavity shift per unit displacement of the collective mode.

In Eq. 2, the collective mode is entirely decoupled from all other ensemble modes due to our approximation of perfectly harmonic and homogeneous trapping. We allow for corrections to this simplified model by introducing a term $H_{\gamma_m}$ representing coupling of the collective mode to a bath, which, in addition to including effects of mixing with other axial modes, might include effects of radial motion or light-induced effects beyond the optomechanical interaction included in $H_{\text{sys}}$. Further accounting for a coupling $H_\kappa$ of the intracavity field to input field modes with energy $H_{\text{drive}}$, the full optomechanical Hamiltonian takes the form

$$H_{\text{tot}} = \hbar \left[ \omega_0 + \delta\omega_N + \mathcal{G} X_0 (\hat{c}^\dagger + \hat{c}) \right] \left( \hat{a}^\dagger \hat{a} - \langle \hat{a}^\dagger \hat{a} \rangle \right) + \hbar \omega_t \hat{c}^\dagger \hat{c} + H_{\text{drive}} + H_\kappa + H_{\gamma_m}, \tag{3}$$

where we have subtracted an offset associated with the average intracavity light level and absorbed the associated force on the atoms into a redefinition of the trap centers $\xi_i$. In terms of the zero-point length $X_0 \equiv \sqrt{\hbar/(2M\omega_t)}$, the annihilation operator $\hat{c}$ for mode $\mathcal{C}$ has the usual definition such that $\hat{X} = X_0(\hat{c}^\dagger + \hat{c})$ and $\hat{P} = iM\omega_t X_0(\hat{c}^\dagger - \hat{c})$.

## II. MECHANICAL MOTION AND TRANSMISSION FLUCTUATIONS

### A. Equations of Motion

Equation 3 is the standard optomechanical Hamiltonian [2, 4] describing a cavity mode, with operator $\hat{a}$, whose frequency is shifted in proportion to the position $X_0(\hat{c}^\dagger + \hat{c})$ of a mechanical oscillator of frequency $\omega_t$. The last three terms in Eq. 3 can be written out explicitly using the standard input-output formalism of quantum optics [5, 6]. One thereby derives equations of motion for $\hat{a}$ and $\hat{c}$ in terms of the optical input operators $\hat{a}_{\text{in}j}$, corresponding to the fields driving the cavity from its two ends labeled by $j \in 1, 2$, and a mechanical input operator $\hat{c}_{\text{in}}$ corresponding to the thermal bath. We shall describe the evolution of the field operators in a rotating frame at the drive frequency $\omega_L$, writing the intracavity field operator $\hat{a} = (\overline{a} + \hat{d})e^{-i\omega_L t}$ in terms of a c-number $\overline{a}$ and a noise operator $\hat{d}$ that accounts for small deviations from the classical value, and similarly letting $\hat{a}_{\text{in}j} = (\overline{a}_{\text{in}j} + \hat{d}_{\text{in}j})e^{-i\omega_L t}$. The classical field values are then in a steady-state relation

$$0 = (i\delta - \frac{\kappa}{2})\overline{a} - \sqrt{\frac{\kappa}{2}}\left(\overline{a}_{\text{in}1} + \overline{a}_{\text{in}2}\right), \tag{4}$$

while the deviations obey the linearized equations of motion [2]

$$\dot{\hat{d}} = (i\delta - \frac{\kappa}{2})\hat{d} + i\alpha\left(\hat{c} + \hat{c}^\dagger\right) - \sqrt{\frac{\kappa}{2}}\left(\hat{d}_{\text{in}1} + \hat{d}_{\text{in}2}\right), \tag{5}$$

$$\dot{\hat{c}} = (-i\omega_t - \frac{\gamma_m}{2})\hat{c} - \sqrt{\gamma_m}\hat{c}_{\text{in}} + i\left(\alpha^*\hat{d} + \alpha\hat{d}^\dagger\right), \tag{6}$$

where $\delta \equiv \omega_L - (\omega_0 + \delta\omega_N)$ represents the detuning of the drive field from cavity resonance for $X = 0$; and $\alpha \equiv -\mathcal{G}X_0\overline{a}$.

Defining the cavity response function $\chi_\kappa(\omega) = 1/[\kappa/2 - i(\omega + \delta)]$ and the mechanical response function $\chi_{\text{m}}(\omega) = 1/[\gamma_m/2 - i(\omega - \omega_t)]$, and rewriting the equations of motion in the Fourier domain, we have:

$$\tilde{d}/\chi_\kappa(\omega) = i\alpha[\tilde{c}(\omega) + \tilde{c}^\dagger(-\omega)] - \sqrt{\frac{\kappa}{2}}[\tilde{d}_{\text{in}1}(\omega) + \tilde{d}_{\text{in}2}(\omega)], \tag{7}$$

$$\tilde{c}/\chi_{\text{m}}(\omega) = i[\alpha^*\tilde{d}(\omega) + \alpha\tilde{d}^\dagger(-\omega)] - \sqrt{\gamma_m}\tilde{c}_{\text{in}}(\omega). \tag{8}$$

Here, $\tilde{\mathcal{O}}(\omega) \equiv \int_{-T/2}^{T/2} e^{i\omega t}\mathcal{O}(t)\,dt/\sqrt{T}$ denotes the windowed Fourier transform [6] of an operator $\mathcal{O}$, and in calculating spectra $\left\langle \tilde{\mathcal{O}}(\omega)\tilde{\mathcal{O}}(-\omega) \right\rangle$ (for Hermitian $\mathcal{O}$) we shall always implicitly take the limit $T \to \infty$ of a long window.

We can solve Eqs. 7-8 to obtain the dependence of $\tilde{X}(\omega)/X_0 = \tilde{c}(\omega) + \tilde{c}^\dagger(-\omega)$ on the optical input fluctuations $\hat{d}_{\text{in}} = (\hat{d}_{\text{in}1} + \hat{d}_{\text{in}2})/\sqrt{2}$ and the mechanical bath operator $\hat{c}_{\text{in}}$:

$$\tilde{c}(\omega) + \tilde{c}^\dagger(-\omega) = \frac{-\sqrt{\gamma_m}[\chi_{\text{m}}^{-1*}(-\omega)\tilde{c}_{\text{in}}(\omega) + \chi_{\text{m}}^{-1}(\omega)\tilde{c}_{\text{in}}^\dagger(-\omega)] - 2\sqrt{\kappa}\omega_t\left[\alpha^*\chi_\kappa(\omega)\tilde{d}_{\text{in}}(\omega) + \alpha\chi_\kappa^*(-\omega)\tilde{d}_{\text{in}}^\dagger(-\omega)\right]}{\mathcal{N}(\omega)}, \tag{9}$$

where

$$\mathcal{N}(\omega) = \chi_{\text{m}}^{-1}(\omega)\chi_{\text{m}}^{-1*}(-\omega) - 2i|\alpha|^2\omega_t\Pi(\omega) \tag{10}$$

and

$$\Pi(\omega) = \chi_\kappa(\omega) - \chi_\kappa^*(-\omega). \tag{11}$$

### B. Input Field

We shall allow one end of the cavity to be driven by a laser at frequency $\omega_L$ that may have some phase noise. To account for this, we let $\hat{d}_{\text{in}1} = \hat{d}_{\text{in}0} + i\beta\overline{a}_{\text{in}1}$, where $\hat{d}_{\text{in}0}$ represents quantum fluctuations, while $\beta(t) \ll 1$ is a real-valued stochastic variable representing the phase noise. At the other end of the cavity, we will admit only vacuum fluctuations (setting $\overline{a}_{\text{in}2} = 0$).

The quantum fluctuations ($j = 0, 2$) satisfy

$$\left\langle \tilde{d}_{\text{in}j}^\dagger(\omega)\tilde{d}_{\text{in}j}(\omega')\right\rangle = 0, \quad \left\langle \tilde{d}_{\text{in}j}(\omega)\tilde{d}_{\text{in}j}^\dagger(\omega')\right\rangle = \delta_T(\omega - \omega'), \tag{12}$$

where in the relevant limit $T \to \infty$, $\delta_T(0) = 1$ and $\delta_T(u) \to 0$ for $u \neq 0$. Phase noise modifies the corresponding relations for $\hat{d}_{\text{in}1}$,

$$\left\langle \tilde{d}_{\text{in}1}^\dagger(\omega)\tilde{d}_{\text{in}1}(\omega')\right\rangle = \left\langle \tilde{\beta}(-\omega)\tilde{\beta}(\omega')\right\rangle |\overline{a}_{\text{in}1}|^2, \quad \left\langle \tilde{d}_{\text{in}1}(\omega)\tilde{d}_{\text{in}1}^\dagger(\omega')\right\rangle = \delta_T(\omega - \omega') + \left\langle \tilde{\beta}(\omega)\tilde{\beta}(-\omega')\right\rangle |\overline{a}_{\text{in}1}|^2, \tag{13}$$

and adds correlations

$$\left\langle \tilde{d}_{\text{in}1}(\omega)\tilde{d}_{\text{in}1}(-\omega')\right\rangle = -\left\langle \tilde{\beta}(\omega)\tilde{\beta}(-\omega')\right\rangle \overline{a}_{\text{in}1}^2. \tag{14}$$

We will parameterize the laser noise by an effective linewidth $\gamma_L(\omega)$ given by the two-sided spectral density of frequency fluctuations, $\gamma_L(\omega) \equiv \omega^2 \left\langle \tilde{\beta}(\omega)\tilde{\beta}(-\omega)\right\rangle$. (For a laser with Lorentzian lineshape, $\gamma_L$ is independent of frequency and represents the full width [7].)

### C. Transmission Spectrum

We now proceed to calculate the two-sided spectrum $S_R^{(2)}(\omega)$ of cavity transmission fluctuations and relate this to the spectrum of the mechanical oscillator's motion. The rate $R$ at which photons are transmitted from the cavity is given in terms of the output field operator $\hat{a}_{\text{out}} = \hat{a}_{\text{in}2} + \sqrt{\frac{\kappa}{2}}\hat{a}$ as $R = \hat{a}_{\text{out}}^\dagger \hat{a}_{\text{out}}$. The fluctuations of this rate about its mean value $\overline{R} = \overline{a}_{\text{out}}^2$ are given by $\hat{a}_{\text{out}}^\dagger \hat{a}_{\text{out}} - \overline{a}_{\text{out}}^2 \approx (\overline{a}_{\text{out}}^* \hat{d}_{\text{out}} + \overline{a}\hat{d}_{\text{out}}^\dagger)$, where $\hat{d}_{\text{out}} = \hat{a}_{\text{out}} - \overline{a}_{\text{out}}$ and we are working to lowest order in $\hat{d}_{\text{out}}/\overline{a}_{\text{out}}$. We assume, without loss of generality, that $\overline{a}$ is real. Defining $\epsilon(\omega) \equiv \tilde{d}_{\text{out}}(\omega) + \tilde{d}_{\text{out}}^\dagger(-\omega)$, we then have $S_R^{(2)}(\omega)/\overline{R} = \langle \epsilon(\omega)\epsilon(-\omega)\rangle$. Using Eq. 7 to evaluate $\tilde{d}_{\text{out}} = \tilde{d}_{\text{in}2} + \sqrt{\kappa/2}\tilde{d}$, we find $\epsilon(\omega) = \epsilon_{\text{opt}}(\omega) + \epsilon_{\text{mech}}(\omega)$, where

$$\epsilon_{\text{opt}}(\omega) = -(\kappa/2)\left[\chi_\kappa(\omega)\tilde{d}_{\text{in}1}(\omega) + \chi_\kappa^*(-\omega)\tilde{d}_{\text{in}1}^\dagger(-\omega)\right]$$
$$+ \left[1 - (\kappa/2)\chi_\kappa(\omega)\right]\tilde{d}_{\text{in}2}(\omega) + \left[1 - (\kappa/2)\chi_\kappa^*(-\omega)\right]\hat{d}_{\text{in}2}^\dagger(-\omega) \tag{15}$$

and

$$\epsilon_{\text{mech}}(\omega) = i\alpha\sqrt{\kappa/2}\Pi(\omega)\left[\tilde{c}(\omega) + \tilde{c}^\dagger(-\omega)\right]. \tag{16}$$

Here, $\epsilon_{\text{opt}}$ contains the intensity fluctuations due to photon shot noise or technical noise of the drive light, whereas $\epsilon_{\text{mech}}$ describes fluctuations in transmission due to atom-induced shifts of the cavity resonance.

Using Eqs. 15 and 16, we can express the fractional fluctuations in transmitted intensity as the sum of three terms describing, respectively, the intrinsic optical fluctuations; the motion-induced fluctuations; and the correlations between the first two:

$$S_R^{(2)}(\omega)/\overline{R}^2 = S_{\text{opt}}^{(2)}(\omega) + S_{\text{mech}}^{(2)}(\omega) + S_{\text{fb}}^{(2)}(\omega). \tag{17}$$

Here,

$$S_{\text{opt}}^{(2)}(\omega) = \langle \epsilon_{\text{opt}}(\omega)\epsilon_{\text{opt}}(-\omega)\rangle/\overline{R} = 1/\overline{R} + \gamma_L(\omega)|\Pi(\omega)|^2 \tag{18}$$

represents the optical fluctuations that would be present even in the absence of optomechanical coupling, namely photon shot noise and laser phase noise (converted into intensity noise by the cavity). $S_{\text{mech}}^{(2)}(\omega)$ is directly related to the spectrum $S_X^{(2)}(\omega) = \left\langle \tilde{X}(\omega)\tilde{X}(-\omega)\right\rangle$ of the mechanical motion (evaluated below in Eq. 21) by

$$S_{\text{mech}}^{(2)}(\omega) = \langle \epsilon_{\text{mech}}(\omega)\epsilon_{\text{mech}}(-\omega)\rangle/\overline{R} = \mathcal{G}^2|\Pi(\omega)|^2 S_X^{(2)}(\omega). \tag{19}$$

Finally, the correlations between the mechanical motion and the optical noise are described by

$$S_{\text{fb}}^{(2)}(\omega) = \langle \epsilon_{\text{opt}}(\omega)\epsilon_{\text{mech}}(-\omega) + \epsilon_{\text{mech}}(\omega)\epsilon_{\text{opt}}(-\omega)\rangle/\overline{R}$$
$$= -4(\mathcal{G}X_0)^2\omega_t\text{Im}\left[\frac{\Pi(\omega)}{\mathcal{N}(\omega)}\left(\chi_\kappa^*(\omega) + 2\gamma_L(\omega)|\Pi(\omega)|^2\overline{R}/\kappa\right)\right]. \tag{20}$$

4## D. Displacement Spectrum

We evaluate the two-sided spectrum of the mechanical oscillator's displacement $S_X^{(2)}(\omega) = \left\langle \tilde{X}(\omega)\tilde{X}(-\omega) \right\rangle$ using Eq. 9. Applying the simplest possible model for the bath, namely quantum white noise [8] with $\left\langle \tilde{c}_{\text{in}}^\dagger(\omega)\tilde{c}_{\text{in}}(\omega) \right\rangle \equiv \langle n_{\text{bath}} \rangle$, we obtain

$$S_X^{(2)}(\omega)/X_0^2 = \frac{\gamma_m \left[ (\langle n_{\text{bath}} \rangle + 1) \left| \chi_m^{-1}(-\omega) \right|^2 + \langle n_{\text{bath}} \rangle \left| \chi_m^{-1}(\omega) \right|^2 \right] + 4\kappa \left| \omega_t \alpha \chi_\kappa(\omega) \right|^2 + 8[\gamma_L(\omega) \overline{R}/\kappa] \left| \omega_t \alpha \Pi(\omega) \right|^2}{|\mathcal{N}(\omega)|^2}. \quad (21)$$

The first pair of terms (in square brackets) describes motion arising from the oscillator's coupling to the bath. The middle term describes motion induced by photon shot noise of the probe light, while the last term describes motion induced by laser frequency noise.

Note that in the absence of optomechanical coupling ($\alpha = 0$), the oscillator spectrum $S_X^{(2)}(\omega)$ reduces to a pair of Lorentzians centered about $\pm \omega_t$

$$\left. \frac{S_X^{(2)}(\omega)}{X_0^2} \right|_{\alpha=0} = \gamma_m \left[ (\langle n_{\text{bath}} \rangle + 1) |\chi_m(\omega)|^2 + \langle n_{\text{bath}} \rangle |\chi_m(-\omega)|^2 \right], \quad (22)$$

whose area is set by the bath temperature

$$\int_{-\infty}^{\infty} \frac{d\omega}{2\pi} \frac{S_X^{(2)}(\omega)}{X_0^2} = 2 \langle n_{\text{bath}} \rangle + 1. \quad (23)$$

## E. Comparison with Measured Spectra

The spectra we measure are one-sided spectra, which we shall denote generically in terms of the two-sided spectra $S^{(2)}(\omega)$ by $S(\omega) \equiv S^{(2)}(\omega) + S^{(2)}(-\omega)$. The actual noise measured at the photodetector, normalized to the average photocurrent $\overline{I}$, is $S_I(\omega)/\overline{I}^2 = S_R(\omega)/\overline{R}^2 + \Sigma_{\text{det}}$. Here, $\Sigma_{\text{det}}$ accounts for imperfect quantum efficiency $Q = 0.5(1)$, an excess noise factor $F = 4.5(5)$ of the avalanche photodiode, and dark (primarily Johnson) noise, and is given by

$$\Sigma_{\text{det}} = \frac{F-Q}{Q} \frac{2}{\overline{R}} + \frac{S_J}{(Q\overline{R})^2}, \quad (24)$$

where $S_J = 1.5(3) \times 10^9 /(\text{s}^2 \text{ Hz})$ expresses the measured dark noise in units of equivalent photon rate. The spectra in Fig. 4 are taken with a photon rate $\overline{R} = 1.2 \times 10^9/\text{s}$ at the output of the cavity, which yields $\Sigma_{\text{det}} = 1.6(1) \times 10^{-8}/\text{Hz}$.

We can now write the measured spectrum as $S_I(\omega)/\overline{I}^2 = S_{\text{mech}}(\omega) + S_{\text{bg}}(\omega)$, where

$$S_{\text{bg}}(\omega) = S_{\text{opt}}(\omega) + S_{\text{fb}}(\omega) + \Sigma_{\text{det}}. \quad (25)$$

From transmission noise $S_{\text{opt}}(\omega)$ measured at large photon rate in the absence of atoms, we have determined the effective laser linewidth to be $\gamma_L(\omega_t) = 2\pi \times 0.8(2)$ kHz at the trap frequency $\omega_t \approx 2\pi \times 500$ kHz (well within our 3 MHz lock bandwidth). At the photon rate $\overline{R} = 1.2 \times 10^9/\text{s}$ used in the spectra of Fig. 4, the associated phase-noise-induced intensity fluctuations are a factor of 1.7(5) below the photon shot noise level; correspondingly, they induce motion (included in our analysis) that is smaller than the zero-point fluctuations $X_0$.

In fitting the measured spectra, we constrain $F$, $Q$, $\gamma_L$ and $\overline{R}$; we leave the dominant background noise contribution $S_J$ free and obtain values consistent with the independently measured dark noise.

## F. Determination of Collective Temperature

Subtracting the background level $S_{\text{bg}}(\omega)$ from the measured spectrum $S_I(\omega)/\overline{I}^2$ allows us to determine $S_X(\omega)$ from Eq. 21 and integrate it to find the occupation of the collective mode:

$$\langle n \rangle + 1/2 = \frac{1}{2} \int_0^\infty \frac{d\omega}{2\pi} S_X(\omega)/X_0^2. \quad (26)$$

5To determine $\langle n \rangle$ in Fig. 2, we use not the spectrum itself but the fractional variance $\sigma_I^2 \equiv \overline{(I-\overline{I})^2}/\overline{I}^2$ of the measured photocurrent $I \propto R$ in a bandwidth $B \gg \omega_t$, which is related to the spectrum $S_I(\omega)/\overline{I}^2$ by

$$\sigma_I^2 = \int_0^B \frac{d\omega}{2\pi} S_I(\omega)/\overline{I}^2 \approx \int_0^\infty \frac{d\omega}{2\pi} S_{\text{mech}}(\omega) + \int_0^B \frac{d\omega}{2\pi} S_{\text{bg}}(\omega). \quad (27)$$

We make in Eq. 19 for $S_{\text{mech}}(\omega)$ the approximation $\Pi(\omega) \approx \Pi(\omega_t) = (2/\kappa)(\mathcal{L}_+ - \mathcal{L}_-)$, which yields

$$S_{\text{mech}}(\omega) \approx (2\mathcal{G}/\kappa)^2 |\mathcal{L}_+ - \mathcal{L}_-|^2 S_X(\omega). \quad (28)$$

Integrating Eq. 28 allows us to obtain $\int \frac{d\omega}{2\pi} S_X(\omega)$ from $\sigma_I^2$ using Eq. 27. The background noise term $\int_0^B \frac{d\omega}{2\pi} S_{\text{bg}}(\omega)$ on the right-hand side of Eq. 27 is independent of the occupation of the collective mode and is well approximated for the data in Fig. 2 by the variance $\sigma_{I,\text{eq}}^2$ measured in the long-time limit. In particular, we can find the change $\langle n \rangle - \langle n \rangle_{\text{eq}}$ in collective mode occupation between two different measurements $\sigma_I^2, \sigma_{I,\text{eq}}^2$ of the fractional transmission variance at fixed background noise by combining Eqs. 26-28:

$$\sigma_I^2 - \sigma_{I,\text{eq}}^2 = 8(\mathcal{G}X_0/\kappa)^2 |\mathcal{L}_+ - \mathcal{L}_-|^2 \left(\langle n \rangle - \langle n \rangle_{\text{eq}}\right). \quad (29)$$

The equilibrium occupation $\langle n \rangle_{\text{eq}} \lesssim 3$ of the collective mode $\mathcal{C}$ is small compared to the values $\langle n \rangle$ plotted in Fig. 2, where mode $\mathcal{C}$ is initially excited. Therefore neglecting $\langle n \rangle_{\text{eq}} \ll \langle n \rangle$, and reexpressing Eq. 29 in terms of transmission rate variances $\sigma^2 \equiv \overline{(R-\overline{R})^2}/\overline{R}^2 = \sigma_I^2 - B\Sigma_{\text{det}}$ and $\sigma_{\text{bg}}^2 \equiv \sigma_{I,\text{eq}}^2 - B\Sigma_{\text{det}}$, we obtain

$$\sigma^2 - \sigma_{bg}^2 = 8(\mathcal{G}X_0/\kappa)^2 |\mathcal{L}_+ - \mathcal{L}_-|^2 \langle n \rangle. \quad (30)$$

### III. ATOM NUMBER CALIBRATION

We measure atom number [9] via the cavity shift $\delta\omega_N = NC\Omega$, where $C = N^{-1} \sum_{i=1}^N \sin^2(k\xi_i)$. Allowing for the small but non-zero radial cloud size $\sigma_r = 7(1)$ $\mu$m $\ll w$, $N$ represents an effective number of on-axis atoms. The cloud is long ($\approx 1$ mm) compared to the 5-$\mu$m beat length between trap and probe, so that $C = 1/2$ in the absence of probe light. Displacement of the atoms by the probe light reduces $C$ by at most 12% in our experiments, and we account for this effect.

### IV. THERMODYNAMIC TEMPERATURE

The thermodynamic axial temperature, given by the mean single-atom vibrational occupation number $\langle n_i \rangle$, is of interest for comparison with the bath temperature inferred from the fits in Fig. 4. We estimate $\langle n_i \rangle$, in an ensemble of $N = 1000(100)$ atoms, by ramping off the trap over 20 $\mu$s $\gg 1/\omega_t$ while increasing the lattice depth of the probe to $U_\text{p} \approx 90\hbar\omega_t$. In the probe lattice, blue-detuned by $\Delta = +280$ MHz from atomic resonance, axially cold atoms localize at positions $x_i'$ near the nodes. Before the cloud has time to expand radially, we determine the atom-probe coupling $C' = \langle \sin^2(kx_i') \rangle$ via the cavity shift, normalized by the shift measured beforehand in the 851-nm lattice at $C = 1/2$. From $C' \approx (\langle n_i' \rangle + 1/2)\sqrt{E_r/U_\text{p}}$ we determine the mean vibrational level $\langle n_i'(0) \rangle = 6(2)$ in the final probe lattice. Note that this measurement provides only an upper bound on $\langle n_i \rangle \leq \langle n_i' \rangle$ if the transfer into the deep probe lattice is not entirely adiabatic.

---

[1] K. W. Murch, K. L. Moore, S. Gupta, and D. M. Stamper-Kurn, Nature Physics **4**, 561 (2008), also see supplementary information.
[2] F. Marquardt, J. P. Chen, A. A. Clerk, and S. M. Girvin, Phys. Rev. Lett. **99**, 093902 (2007).
[3] P. Rabl, C. Genes, K. Hammerer, and M. Aspelmeyer, Phys. Rev. A **80**, 063819 (2009).
[4] I. Wilson-Rae, N. Nooshi, W. Zwerger, and T. J. Kippenberg, Phys. Rev. Lett. **99**, 093901 (2007).
[5] C. W. Gardiner and M. J. Collett, Phys. Rev. A **31**, 3761 (1985).
[6] A. A. Clerk, M. H. Devoret, S. M. Girvin, F. Marquardt, and R. J. Schoelkopf, Rev. Mod. Phys. **82**, 1155 (2010), also see online appendices at http://rmp.aps.org/epaps/RMP/v82/i2/p1155_1/QNoiseRMPAppsApr2010.pdf.

[7] G. Audoin, in *Metrology and Fundamental Constants, corso 68 of the International School of Physics Enrico Fermi*, edited by P. G. A. F. Milone and F. Leschiutta (North-Holland, Amsterdam, 1980), pp. 169–222.
[8] C. W. Gardiner and P. Zoller, *Quantum Noise* (Springer, Berlin, 2000).
[9] M. H. Schleier-Smith, I. D. Leroux, and V. Vuletić, Phys. Rev. Lett. **104**, 073604 (2010).


\end{document}